\documentclass[review]{elsarticle}

\usepackage{lineno,graphicx,caption,subcaption,amsmath}

\begin{document}
\begin{frontmatter}

\title{A method for characterizing after-pulsing and dark noise of PMTs and SiPMs }

\author[RHUL]{A. Butcher}
\author[TRIUMF]{L. Doria}
\author[RHUL]{J. Monroe}
\author[TRIUMF]{F. Reti\`{e}re \corref{corr}}
\author[TRIUMF]{B. Smith}
\author[RHUL]{J. Walding}


\address[RHUL]{Royal Holloway, University of London, Egham Hill, Egham, Surrey TW20 0EX, United Kingdom}
\address[TRIUMF]{TRIUMF, Vancouver, British Columbia, V6T 2A3, Canada}
\cortext[corr]{Corresponding author, fretiere@triumf.ca}

\begin{abstract}
Photo-multiplier tubes (PMTs) and silicon photo-multipliers (SiPMs) are detectors sensitive to single photons that are widely used for the detection of scintillation and Cerenkov light in subatomic physics and medical imaging. This paper presents a method for characterizing two of the main noise sources that PMTs and SiPMs share: dark noise and correlated noise (after-pulsing). The proposed method allows for a model-independent measurement of the after-pulsing timing distribution and dark noise rate. 
\end{abstract}

\end{frontmatter}


\section{Introduction}

Single-photon sensitive detectors include a large gain stage for converting the single charge carrier produced by photon absorption into detectable electrical pulses.  To preserve stable gain from event to event, it is critical to prevent the charge carriers created during the amplification process from themselves generating additional pulses. In practice, the generation of pulses correlated with earlier pulses is impossible to completely avoid and significant effort goes into minimizing it. Then, the analysis employing the photo-detector must cope appropriately with such correlated delayed pulses (CDP). In this paper, we introduce the CDP acronym because the meaning and origin of after-pulsing, the commonly used term, changes with the type of photo-detector being studied.  In this paper, we show that the same method can be used for characterizing the CDP distributions of the most commonly used single-photon detectors: photo-multiplier tubes (PMTs) and silicon photo-multipliers (SiPMs).	

In PMTs, CDPs arise from at least two separate processes: the inelastic scattering of electrons on the first dynode and the production of ions by collision of electrons with residual gas atoms~\cite{morton1967afterpulses,coates1973}. The former process is often called double pulsing, and it typically occurs over time scales smaller than 100ns. Double pulsing  is generated by an electron from the photo-cathode striking the first dynode inelastically (at least) twice. The latter process is labeled after-pulsing and occurs over $\mu$s time scales (depending on the size of the PMT) being driven by the velocity of the ions drifting back from their production point---most likely in the inter-dynode space---back to the photo-cathode.  Both processes can compromise the performance of experiments relying on pulse shape discrimination for identifying nuclear and electronic recoils in scintillating materials. Such experiments must characterize CDPs with care in order to achieve optimum performance~\cite{DEAP1}.

In SiPMs, CDPs arise from at least two processes: the production of photons during the avalanche in the gain amplification stage, and the trapping and subsequent release of charge carriers produced in avalanches~\cite{van2010comprehensive}. The latter process is usually labeled after-pulsing. The former process is usually called cross-talk and it is subdivided in two categories: prompt and delayed cross-talk. This physics-process-driven terminology is somewhat confusing as delayed cross-talk may look exactly like after-pulsing. In this paper, we introduce a method for characterizing correlated delayed pulses regardless of their physical origin. Understanding the physical origin of CDPs is important because they can limit the performances of SiPMs by preventing operation at high enough voltage, but it is also highly desirable to measure their rate and timing distribution without having to make any a priori model assumption. 

Accurate measurement of dark noise, excluding any contribution from CDPs, is also important for understanding the timing distribution of incoming photons. The method proposed in this paper excludes CDPs from the dark-noise measurement, so long as the dark noise rate does not overwhelm the CDP rate.

\section{Method}
\label{sec:method}

The method is based on using the time distribution between two consecutive pulses.  Selection of the pulse at $t = 0$, which we will call the primary pulse, must be done with care. Primary pulses can arise either from dark-noise or be generated by single photons using a very faint light source. With a light source the photon count distribution follows a Poisson distribution in the light intensity range of interest, and pulses generated by a single photon are typically identified by selecting a specific charge range. Selecting single photo-electrons is relatively straightforward with SiPMs thanks to their small gain fluctuations. For PMTs, gain fluctuations do not allow an unambiguous selection of single photo-electrons. In this case the most accurate method is to operate at very low light intensity, ensuring negligible probability of producing more than one photo-electron within a certain narrow time window. 

We will call the pulse immediately following the primary pulse the secondary pulse. No specific charge selection is necessary for this pulse, except for ensuring that it is not electronics noise. When selecting events to be used in the analysis, it is important to ensure that CDPs not associated with the selected primary pulse have a negligible chance to be selected as the secondary pulse.  In practice, this is usually achieved by only using events in which there is no other pulse in a significant time window before the primary pulse. The exclusion time window must be large enough that the probability of a pulse at $t>0$ being a CDP from a pulse at $t<0$ is negligible.

The measured distribution of time differences between secondary and primary pulses gives the probability that the secondary pulse occurs at time $t$ following a primary pulse at time $t=0$. In practice the measured distribution we employ here is a histogram with the content of each bin, $p_i$, being the probability per unit time that a secondary pulse occurs within the bin limits, $t_i$ and $t_{i+1}$. By construction,  the probability must integrate to 1 between $t=0$ and $t=\infty$. With $N$ as the total number of events and $n_i$ as the number of events with a secondary pulse between $t_i$ and $t_{i+1}$ then
\begin{equation}
\label{eqn:psec}
p_i=\frac{n_i}{N (t_{i+1}-t_i)}
\end{equation}
and the statistical error is
\begin{equation}
\label{eqn:psecerr}
\sigma_{p_i}=\frac{\sqrt{n_i}}{N (t_{i+1}-t_i)}
\end{equation}
The rationale for using the timing distribution of the secondary pulse and not the timing distribution of all the pulses following the primary pulse is to avoid correlated pulses of correlated pulses, which are very difficult to deal with statistically.  The drawback of using the timing distribution of the secondary pulse is that early processes shadow late processes. In the past, this issue was dealt with by fitting the $p_i$ distribution by making some model assumption about after-pulsing in order to account for the shadowing effect ~\cite{Vacheret201169,Rosado2015153,piemonte2012development,Du2008396}. What follows shows that this is not necessary.

The probability $P_i$ that the secondary pulse occurs between $t_i$ and $t_{i+1}$ can be written using Poisson statistics with $\beta_i$ as the average number of correlated pulses occurring before $t_i$, and $\lambda_i$ as the average number of correlated pulses occurring between $t_i$ and $t_{i+1}$, as
\begin{equation}
\label{eqn:psecshadow}
P_i =p_i (t_{i+1}-t_i)= e^{-\beta_i} \left(1-e^{-\lambda_i}\right).
\end{equation}
The first term is the probability that no pulse occurs prior to $t_i$ and the second term is the probability that at least one pulse occurs between  $t_i$ and $t_{i+1}$. It is possible to extract the distribution of $\lambda_i$ experimentally by inverting equation~\ref{eqn:psecshadow} without making any model assumptions, realizing that:
\begin{equation}
\label{eqn:SumPrev}
\beta_i = \sum_{j=0}^{i-1}\lambda_j,
\end{equation}
with $\beta_0=0$ and
\begin{equation}
\label{eqn:unshadowMean}
\lambda_i=-\ln\left(1-\frac{P_i}{e^{-\beta_i}}\right)
\end{equation}
So, one can evaluate $\lambda_i$ starting with $\lambda_0=0$ and building the sum $\beta_i$. Again $\lambda_i$ is the average number of secondary pulses occurring between $t_i$ and $t_{i+1}$. It should then be normalized per unit time to get the pulse rate $\rho_i$ as
\begin{equation}
\rho_i=\frac{\lambda_i}{t_{i+1}-t_{i}}.
\end{equation}
The rate of correlated delayed pulses is expected to vanish at sufficiently large time, leaving only the dark noise rate $R_\mathrm{DN}$. Error propagation yields
\begin{equation}
\label{eqn:unshadowSig}
\sigma_{\lambda_i}=\frac{\sqrt{\sigma_{P_i}^2+P_i^2\sigma_{\beta_i}^2}}{e^{-\beta_i}-P_i},
\end{equation}
with $\sigma_{\beta_i}$ the error on the mean number of pulses occurring prior to $t_i$. Recalling that $\beta_i =\beta_{i-1} + \lambda_{i-1}$ and $\sigma_{\beta_0}=0$ , and assuming that the errors on $\beta_{i-1}$ and $p_{i-1}$ are independent yields
\begin{equation}
\label{eqn:unshadowSigSum}
\sigma_{\beta_i}=\frac{\sqrt{\sigma_{P_{i-1}}^2+(e^{-\beta_{i-1}}\sigma_{\beta_{i-1}})^2}}{e^{-\beta_{i-1}}-P_{i-1}}.
\end{equation}
This formula can be used for calculating the errors starting from bin 0. 

In summary, the procedure to characterize dark noise and CDP is:
\begin{enumerate}
\item select suitable primary pulses usually corresponding to single photo-electrons
\item build the time distribution between the primary and secondary pulse
\item calculate the CDP plus dark noise rate, $\rho_i$; at long times this is simply the dark noise rate, $R_\mathrm{DN}$
\item subtract $R_\mathrm{DN}$ from $\rho_i$ to obtain the CDP timing distribution and rate.
\end{enumerate}

The key assumption of the method is that all pulses are identified with high efficiency, otherwise the timing distribution may be distorted.  In particular, primary and secondary pulses may be hard to disentangle when the secondary pulse occurs shortly after the primary pulse. In this case one solution is to merge primary and secondary pulses when calculating the primary pulse charge, so that these events are eliminated by the charge selection criteria on the primary pulse. Doing so implicitly assumes that short time scales can be safely neglected for the application of interest. This method also does not work in the presence of deadtime that would blank a certain timing window. Dark-noise also limits the reach of the CDP measurement to time scales not much longer than the inverse of the dark-noise rate. 

For illustration, we will now apply this method to two simulated examples representing the PMT and SiPM cases, and one data set for SIPMs. 

\section{Example applications for PMTs and SiPMs based on simulations}
\label{sec:examples}

Two examples have been simulated using a simple Monte Carlo routine generating single photo-electron pulses at time $t=0$, followed by correlated delayed pulses and dark noise:
\begin{itemize}
\item a PMT with a dark-noise rate of 1kHz and a Gaussian delayed pulse distribution of mean $6 \mu\mathrm{s}$ and width $1 \mu\mathrm{s}$. This  example is a simplified model of the large area PMTs used in various experiments---for example Hamamatsu R5912 PMTs used in the Daya Bay~\cite{DayaBayPMT} and DEAP-3600~\cite{DEAPPMT} experiments or Hamamatsu R7081 used in the Double Chooz experiment~\cite{Ap2DChooz}. 
\item a SiPM with a dark noise rate of 100kHz and an exponential delayed pulse distribution with time constant $100 \mathrm{ns}$. This is again a simplified model of after-pulsing~\cite{van2010comprehensive}. Recovery of the SiPM is neglected.
\end{itemize}

The number of dark noise pulses in each throw is Poisson-distributed, and their times are uniformly-distributed between $t=0$ and $t=100 \mu\mathrm{s}$. Dark noise pulses at negative times are not simulated, as the analysis method requires that there are no pulses for a significant time before $t=0$. The delayed pulse chain is recursively simulated for each dark noise pulse and the primary pulse, following the properties described above. Separate simulations are performed with the mean number of delayed pulses, $\langle N_\mathrm{CDP}\rangle$, being 0.1, 0.2 and 0.5. The probability and timing distribution of producing a delayed pulse from a delayed pulse is assumed to be the same as for a delayed pulse from a primary pulse or dark noise pulse. This latter assumption is a simplification. In PMTs, after-pulses generated by ions hitting the photo-cathode are known to yield on average more than one photo-electron. In SiPMs, after-pulses can generate more than one avalanche due to prompt cross-talk. This simplification, however, is not important because the timing distribution of all the pulses is only shown for demonstrating the method and it is not used in the analysis. 

\begin{figure}
\centering
\begin{subfigure}{.5\textwidth}
  \centering
  \includegraphics[width=\linewidth]{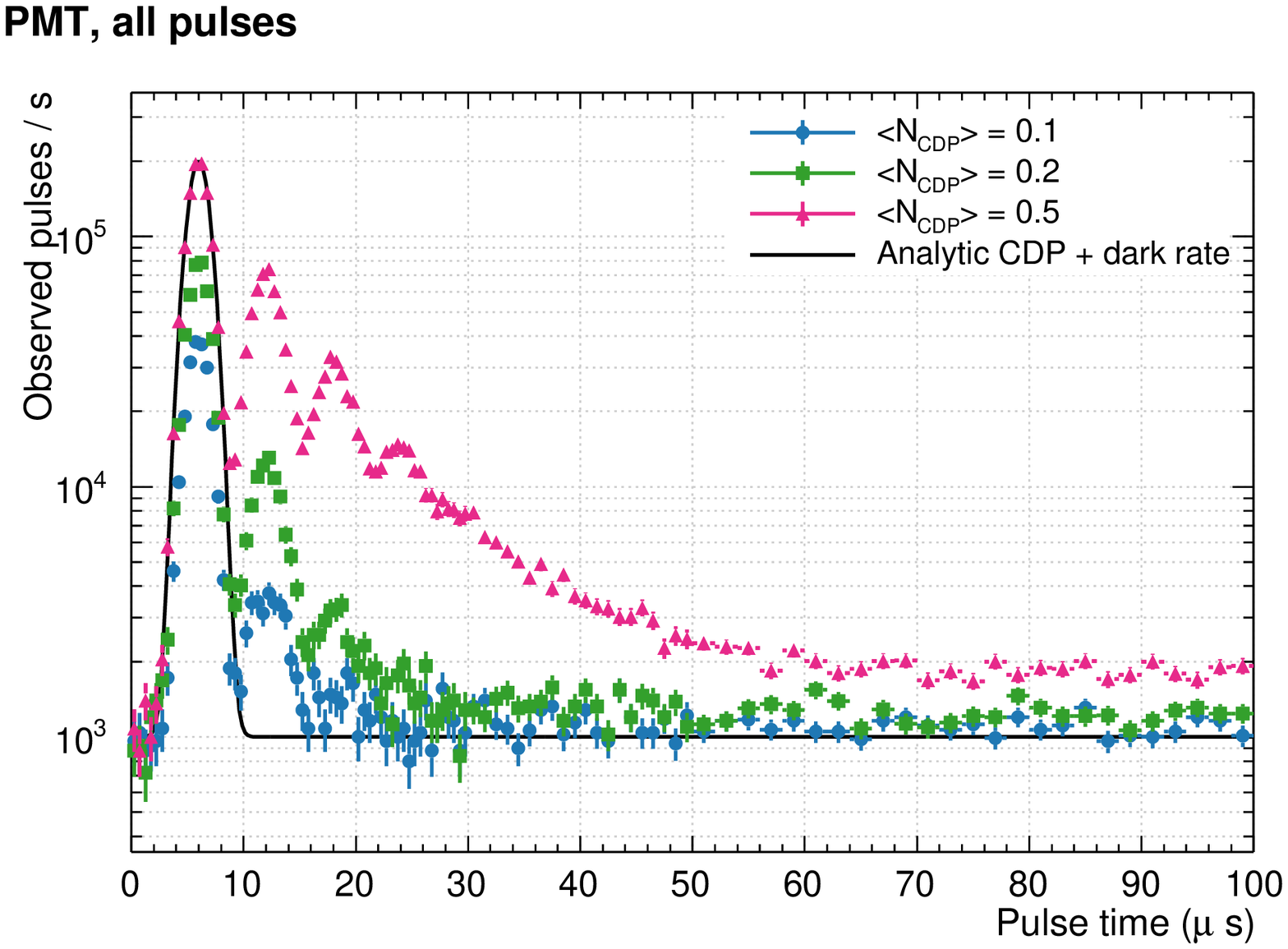}
\end{subfigure}%
\begin{subfigure}{.5\textwidth}
  \centering
  \includegraphics[width=\linewidth]{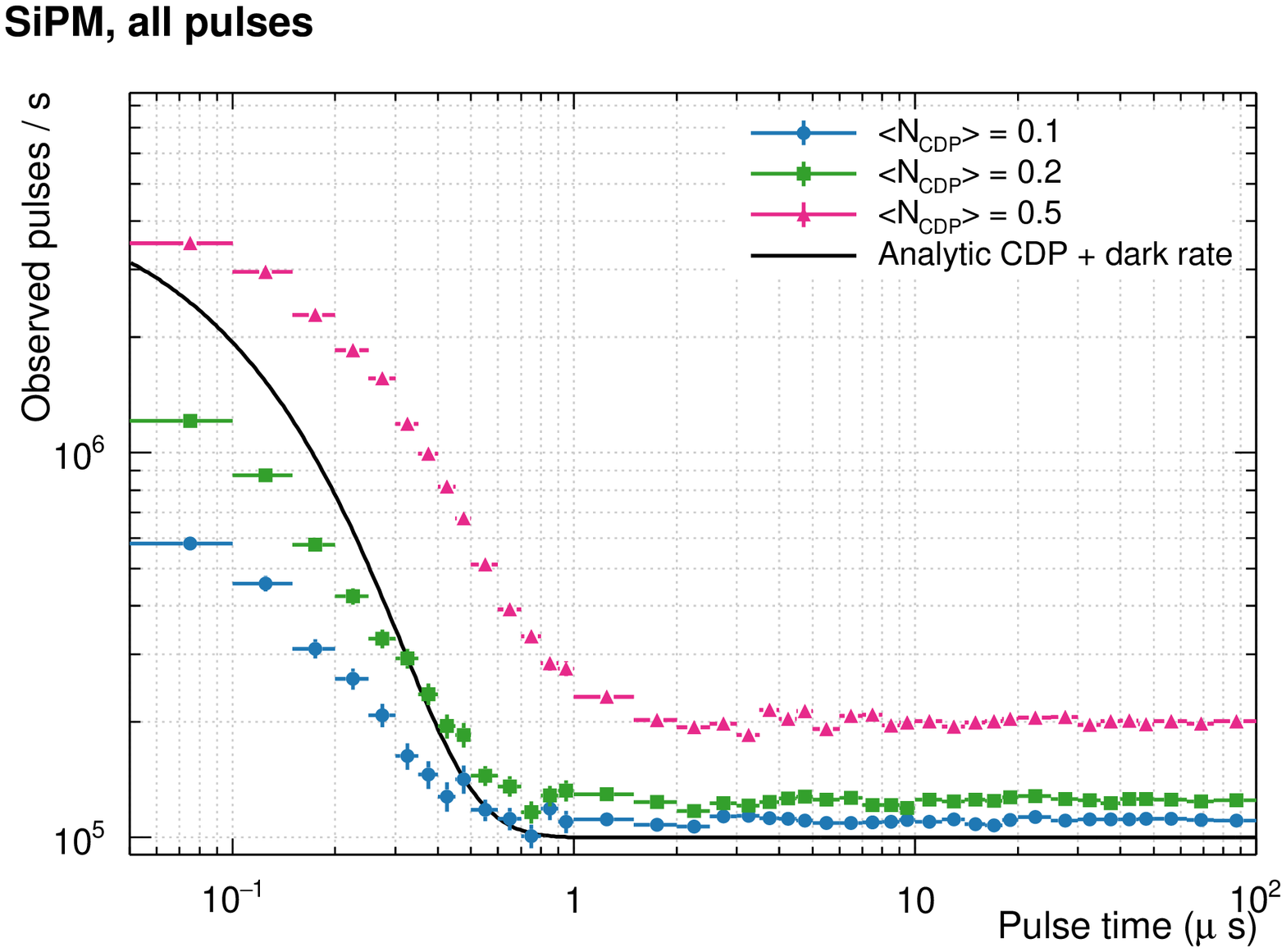}
\end{subfigure}
\begin{subfigure}{.5\textwidth}
  \centering
  \includegraphics[width=\linewidth]{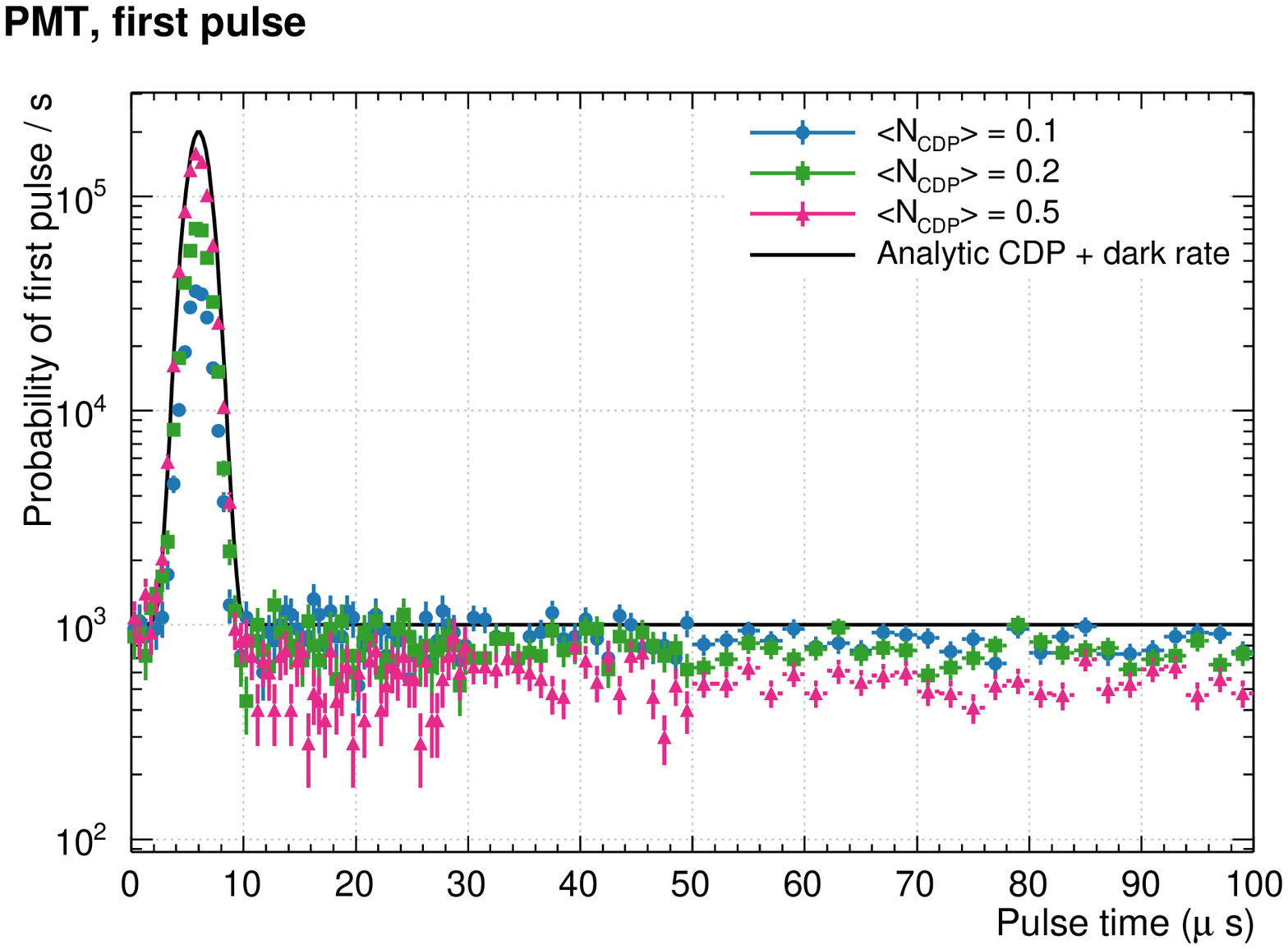}
\end{subfigure}%
\begin{subfigure}{.5\textwidth}
  \centering
  \includegraphics[width=\linewidth]{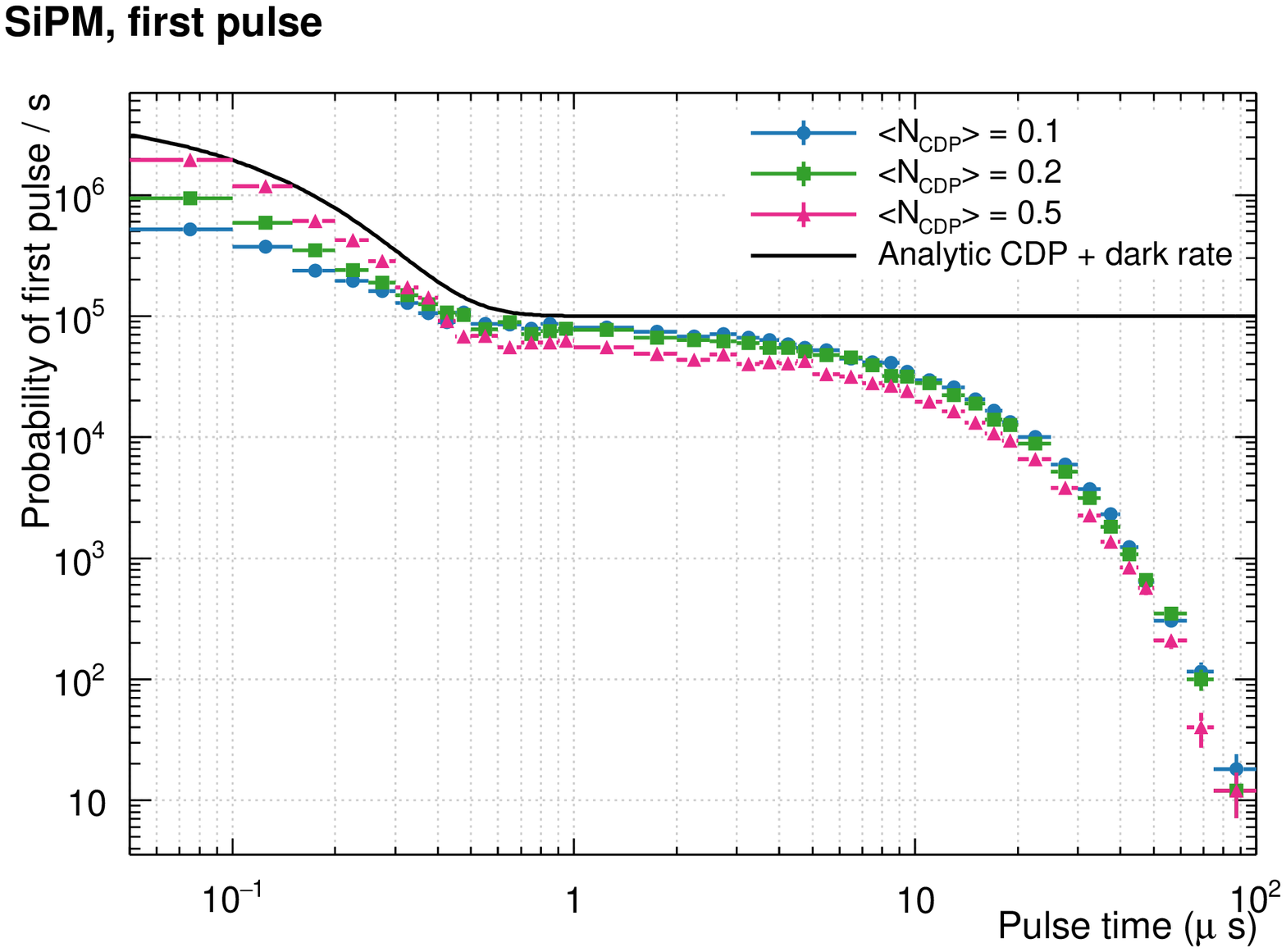}
\end{subfigure}
\begin{subfigure}{.5\textwidth}
  \centering
  \includegraphics[width=\linewidth]{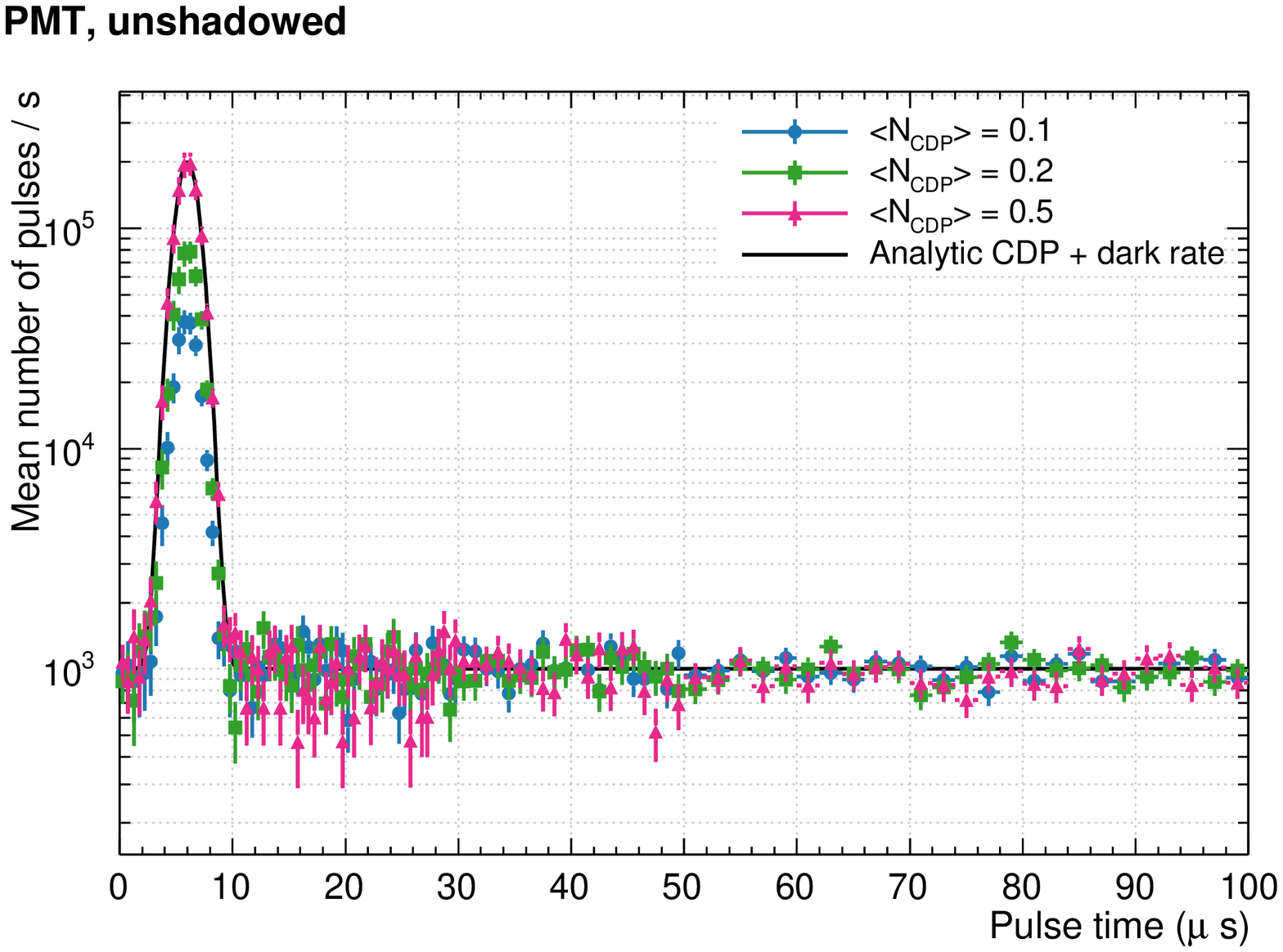}
\end{subfigure}%
\begin{subfigure}{.5\textwidth}
  \centering
  \includegraphics[width=\linewidth]{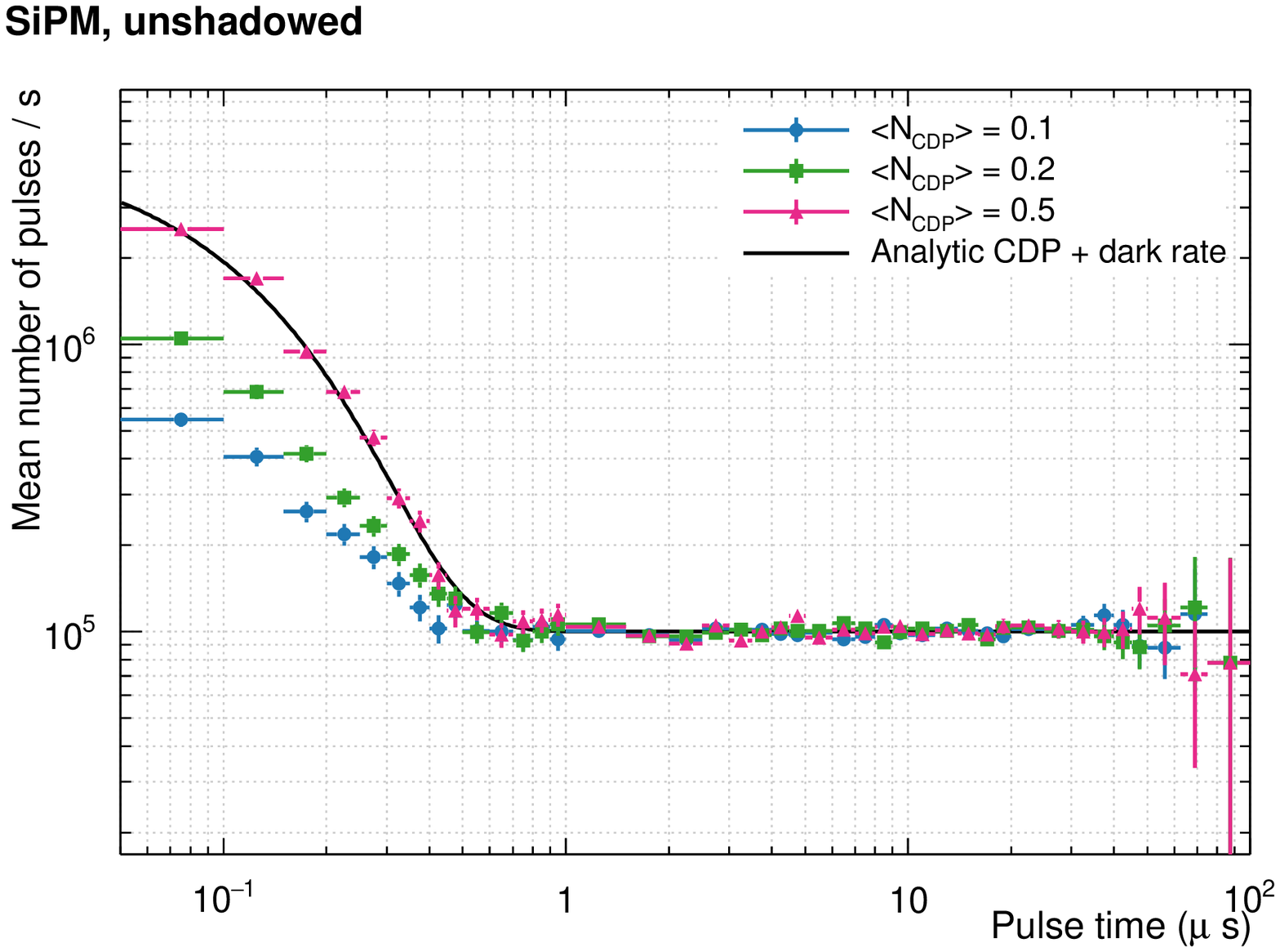}
\end{subfigure}
\caption{Observed (top), first pulse (middle) and unshadowed (bottom) time distributions for PMT (left) and SiPM (right) simulations, as described in the text. The blue, green and magenta points represent simulations with mean numbers of correlated delayed pulses of 0.1, 0.2 and 0.5, respectively. The black lines show the theoretical distribution of the dark noise and delayed pulse spectra for a mean number of delayed pulses of 0.5. For the PMT case the theoretical distribution is $1000+10^6\times0.5\times\mathrm{Gaussian}(6,1)$. For the SiPM case the theoretical distribution is $1000+10^6\times0.5\times\mathrm{Exp}(-t/0.1)$. The factor of $10^6$ accounts for the x-axis using microseconds and the y-axis using seconds.}
\label{fig:pulses}
\end{figure}

The top row of figure~\ref{fig:pulses} shows the pulse time distributions of all the pulses that occurred after $t=0$. The solid lines in the plots show the distributions that would be expected if there were only dark noise pulses and delayed pulses of the primary pulse---i.e.\ if there were no delayed pulses from delayed pulses, and no delayed pulses from dark noise pulses.

The middle row of figure~\ref{fig:pulses} shows the time distribution of the secondary pulse, which is defined as the first pulse following the primary pulse. As noted, the integral of this distribution is 1 by construction.

The bottom row of figure~\ref{fig:pulses} shows the time distributions after applying the unshadowing procedure described in section~\ref{sec:method}. The y-axis of these plots is the rate of pulses at a given time, and the dark noise rate can be read as the value on the y-axis in the latter flat part of the spectrum, assuming  that the rate of correlated delayed pulses of the primary pulse becomes negligible compared to the dark noise rate at long times. 

Finally, we fit a horizontal straight line to the flat parts of each unshadowed distribution to extract $R_\mathrm{DN}$. We fit the range 10--100$\mu$s for the PMT case, and 2--100$\mu$s for the SiPM case. We then subtract $R_\mathrm{DN}$ from the unshadowed distributions, and integrate to find $\langle N_\mathrm{CDP}\rangle$ without making any model assumption regarding the after-pulsing timing distribution. The results of this procedure are shown in table \ref{table:fit}. The extracted values agree with the simulated values within uncertainty.

\begin{table}[h]
  \centering
  \begin{tabular}{l | l | l | l }
      Simulated $R_\mathrm{DN}$ & Fitted $R_\mathrm{DN}$ & Simulated $\langle N_\mathrm{CDP}\rangle$ & Computed $\langle N_\mathrm{CDP}\rangle$\\ \hline 
      1~kHz & $(0.97\pm0.02)$~kHz & 0.1 & $0.101\pm0.005$ \\
      1~kHz & $(0.99\pm0.02)$~kHz & 0.2 & $0.20\pm0.01$ \\
      1~kHz & $(0.97\pm0.02)$~kHz & 0.5 & $0.51\pm0.02$ \\ \hline
      100~kHz & $(100\pm1)$~kHz & 0.1 & $0.093\pm0.006$ \\
      100~kHz & $(99\pm1)$~kHz & 0.2 & $0.201\pm0.006$ \\
      100~kHz & $(100\pm1)$~kHz & 0.5 & $0.495\pm0.009$ \\
     
  \end{tabular}
    \caption{Simulated and extracted parameters for the unshadowed distributions shown in figure~\ref{fig:pulses}.}
  \label{table:fit}
\end{table}

\section{Example of application to SiPM data}
\begin{figure}
\centering
\begin{subfigure}{.5\textwidth}
  \centering
  \includegraphics[width=\linewidth]{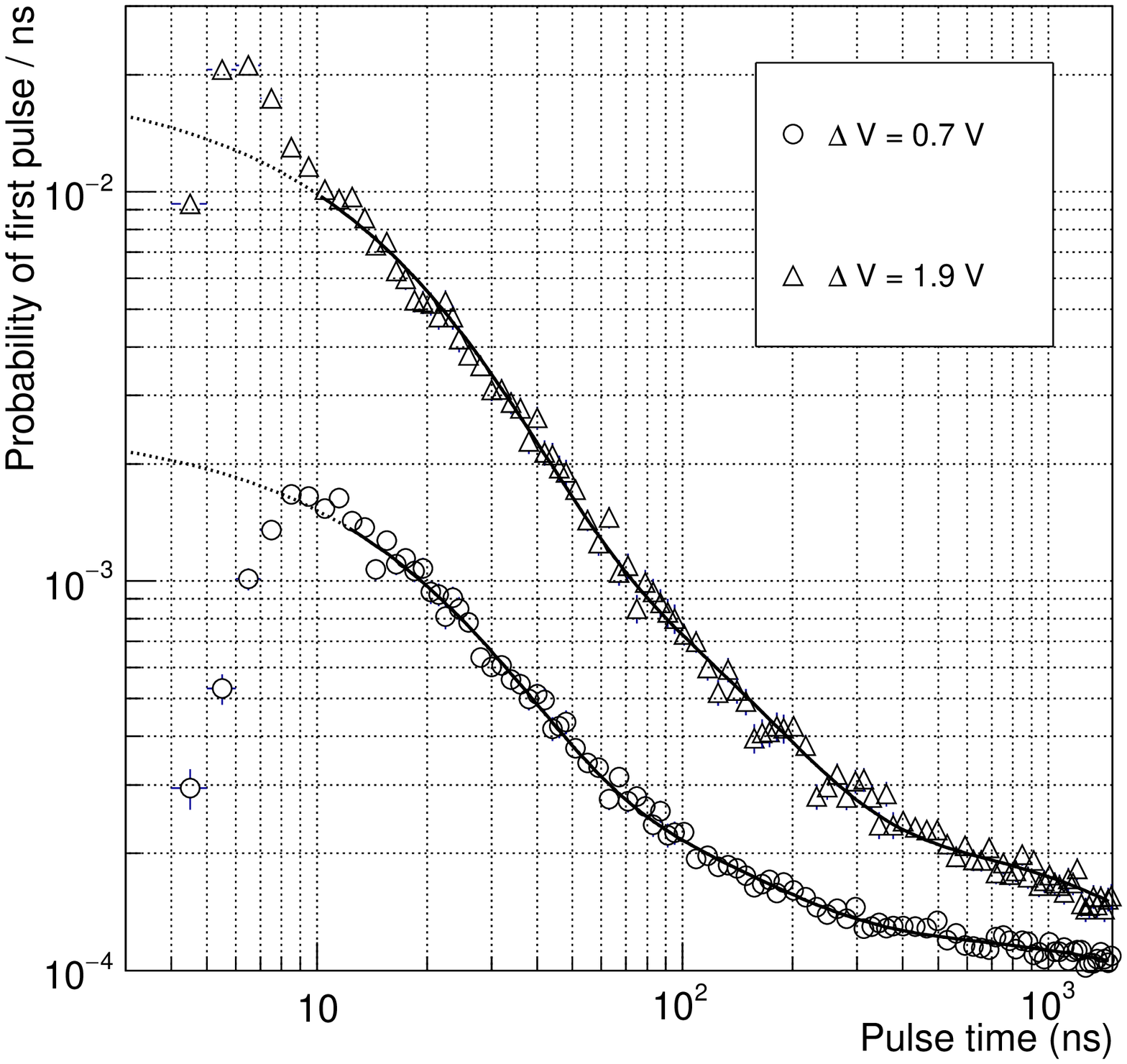}
\end{subfigure}%
\begin{subfigure}{.5\textwidth}
  \centering
  \includegraphics[width=\linewidth]{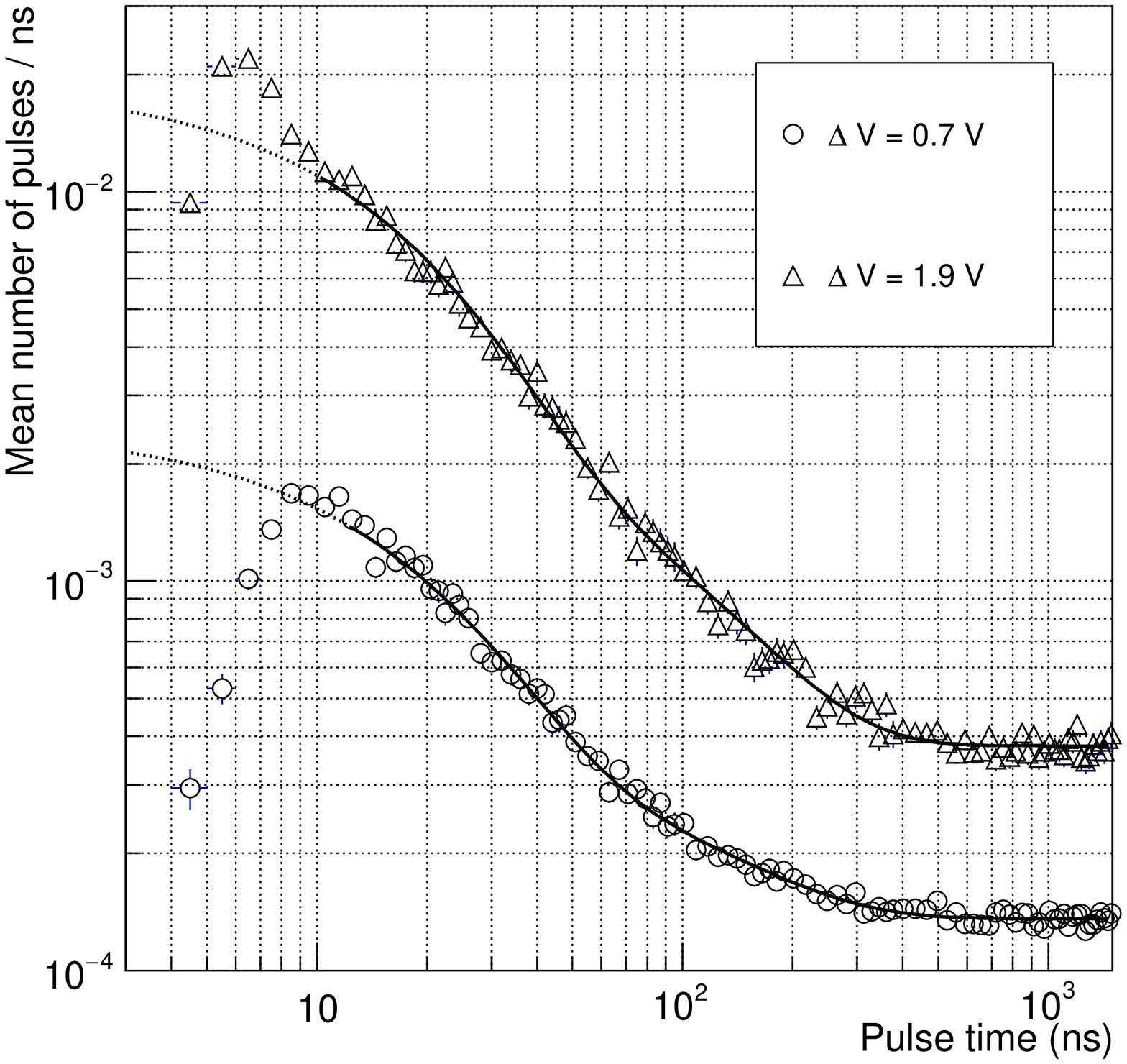}
\end{subfigure}
\caption{Hamamatsu MPPC data from \cite{Du2008396} using the first pulse method on the left and the new unshadowed method on the right}
\label{fig:Data}
\end{figure}

In order to further support the proposed method, previously published data in~\cite{Du2008396} are re-analyzed converting the probability of the next avalanche distribution into the avalanche rate as shown in Figure ~\ref{fig:Data} left-hand and right-hand sides respectively. The data are for Hamamatsu $1 \times 1~\mathrm{mm}^2$ Multi-Pixel Photon Counter produced in 2006 and operated at room temperature. Two data sets are used, one with low after-pulsing at low over-voltage (0.7V) and one with large after-pulsing at high over-voltage (1.9V). 

 The total pulse rate is calculated by integrating the rate histogram over the full range of the data, which in this case is 1.5~$\mu$s. The average number of pulses occurring within this window following a single avalanche pulse is 0.252$\pm$0.001 and 0.96$\pm$0.01 at 0.7~V and 1.9~V over-voltage respectively. The dark noise contribution can be estimated looking at the data beyond 500~ns. The dark noise rate is measured by fitting a constant trough the data above 500ns, yielding 135.9 $\pm$ 0.9 kHz and 377 $\pm$ 4 kHz at 0.7~V and 1.9~V over-voltage respectively. Then, the average number of dark noise pulses occurring within the 1.5 $\mu$s can be calculated and subtracted from the total number of pulses, yielding the after-pulsing rate, i.e. the average number of after-pulses triggered by a single avalanche. It is 0.047 $\pm$ 0.01 and 0.40 $\pm$ 0.01 at 0.7~V and 1.9~V over-voltage respectively. 
 
 In the original publication, dark noise and after-pulsing were assessed by fitting the first pulse probability distribution to a model assuming two after-pulsing time constants. The same model can be applied to the rate data as follow:
\begin{equation}
\label{eqn:RateTwoAP}
R(t) = \frac{\alpha_S  e^{-t/\tau_S}}{\tau_S} + \frac{\alpha_L e^{-t/\tau_L}}{\tau_L}+R_\mathrm{DN}
\end{equation}
With $\alpha_S$ and $\alpha_L$ the average number of after-pulses per parent avalanche for the long and short after-pulsing respectively and $\tau_S$ and $\tau_L$ the short and long after-pulsing time constants respectively and $R_{DN}$ the dark noise rate. This equation assumes that after-pulsing is driven by Poisson statistics, hence more than one after-pulse can occur in principle.  On the other hand, equation in ~\cite{Du2008396} assumed that one avalanche can generate at most one after-pulse. Using Poisson statistics is more justified for SiPM, hence we introduced a new function for fitting the first pulse probability distribution as follow:
\begin{equation}
\label{eqn:firtPulseProbData}
p_i = e^{-\beta_i} (\frac{\alpha_S  e^{-t_i/\tau_S}}{\tau_S} + \frac{\alpha_L e^{-t_i/\tau_L}}{\tau_L}+R_\mathrm{DN})
\end{equation}
and
\begin{equation}
\label{eqn:SumPrevData}
\beta_i = \int_{t_0}^{t_i}(\frac{\alpha_S  e^{-t/\tau_S}}{\tau_S} + \frac{\alpha_L e^{-t/\tau_L}}{\tau_L} + R_\mathrm{DN}) dt
\end{equation}
This function is valid provided that the probability within each bin is small, otherwise the second term of equation \ref{eqn:firtPulseProbData} should use the Poisson formalism.  The fit excludes data points below $t_0$=10~ns because the fit function does not account for the suppression of after-pulsing due to the SiPM recovery. To get consistent results between the fits of the rate and first pulse distributions, the first pulse probability distribution must be normalized excluding the early data points and the same normalization must be accounted for in the fit function.  The total after-pulsing rates obtained by fitting are 0.060 $\pm$ 0.002 and 0.444 $\pm$ 0.012 at 0.7~V and 1.9~V over-voltage respectively, which are larger than the numbers obtained by counting. The difference is due to the incorrect extrapolation of the fit function below 10~ns that do not account for the suppression of after-pulsing due to recovering. The estimates obtained by counting are  the correct ones. This example clearly demonstrates the advantage of assessing after-pulsing in a model independent fashion without having to rely on a functional form that can be very difficult to define accurately.

\section{Conclusion}

\label{sec:conclusion}

This paper describes a method for characterizing dark noise and correlated delayed pulse distributions that does not rely on modeling the delayed pulse distribution (for example as a Gaussian or an exponential). The method yields a histogram representing the rate of a dark noise pulses and correlated delayed pulses (CDPs) occurring as a function of time. It removes the complications associated with modeling CDPs of CDPs, and CDPs of dark noise pulses. We have demonstrated the method on Monte Carlo simulations of PMTs and SiPMs, and shown that the method extracts the expected dark noise and delayed pulse distributions.  Subtracting dark noise yields the rate of CDPs as a function of time, which can be readily used in the simulation of an experiment; this can be critical for accurate reconstruction of event position and event pulse shape discrimination variables.

We have also applied this method to previously published data showing that the after-pulsing rate had been over-estimated because it had been assessed using a function that did not reproduce the data well at short time scales (under 10~ns). The proposed method is more accurate because it does not rely on any a priori knowledge of the  CDP rate distribution. Overall the proposed method significantly ease the interpretation of the CDP rate hence enabling precision characterization of the various sources of photo-detector CDPs:  electron and ion backflow for PMTs, carrier trapping and delayed cross-talk for SiPMs.

\section*{Acknowledgements}
\label{sec:ack}

This work is supported by the National Science and Engineering Research Council of Canada (NSERC), the European Research Council (ERC StG 279980) and the Leverhulme Trust (ECF-20130496). We are grateful to Chris Jillings from SNOLAB for useful discussions and comments about the manuscript.


\section*{References}


\begin{thebibliography}{10}

\bibitem{morton1967afterpulses}
G.~Morton, H.~Smith, R.~Wasserman, Afterpulses in photomultipliers, IEEE
  Transactions on Nuclear Science 14~(1) (1967) 443--448.

\bibitem{coates1973}
P.~Coates, 
The origins of
  afterpulses in photomultipliers, J. Phys. D 6~(10) (1973) 1159.

\bibitem{DEAP1}
P.-A. Amaudruz, et~al.,
  of the scintillation time spectra and pulse-shape discrimination of
  low-energy $\beta$ and nuclear recoils in liquid argon with {DEAP-1},
  Astroparticle Physics 85 (2016) 1 -- 23.

\bibitem{van2010comprehensive}
H.~T. Van~Dam, S.~Seifert, R.~Vinke, P.~Dendooven, H.~Lohner, F.~J. Beekman,
  D.~R. Schaart, A comprehensive model of the response of silicon
  photomultipliers, IEEE Transactions on Nuclear Science 57~(4) (2010)
  2254--2266.

\bibitem{Vacheret201169}
A.~Vacheret, et~al.,
  and simulation of the response of multi-pixel photon counters to low light
  levels, Nuclear Instruments and Methods in Physics Research Section A:
  Accelerators, Spectrometers, Detectors and Associated Equipment 656~(1)
  (2011) 69 -- 83.

\bibitem{Rosado2015153}
J.~Rosado, V.~Aranda, F.~Blanco, F.~Arqueros,
  Modeling
  crosstalk and afterpulsing in silicon photomultipliers, Nuclear Instruments
  and Methods in Physics Research Section A: Accelerators, Spectrometers,
  Detectors and Associated Equipment 787 (2015) 153 -- 156, new Developments in
  Photodetection \{NDIP14\}.

\bibitem{piemonte2012development}
C.~Piemonte, A.~Ferri, A.~Gola, A.~Picciotto, T.~Pro, N.~Serra, A.~Tarolli,
  N.~Zorzi, Development of an automatic procedure for the characterization of
  silicon photomultipliers, Nuclear Science Symposium and Medical Imaging
  Conference (NSS/MIC), IEEE (2012) 428--432.

\bibitem{Du2008396}
Y.~Du, F.~Reti\`{e}re,
 After-pulsing
  and cross-talk in multi-pixel photon counters, Nuclear Instruments and
  Methods in Physics Research Section A: Accelerators, Spectrometers, Detectors
  and Associated Equipment 596~(3) (2008) 396 -- 401.

\bibitem{DayaBayPMT}
S.~Jetter, D.~Dwyer, J.~Wen-Qi, L.~Da-Wei, W.~Yi-Fang, W.~Zhi-Min,
  W.~Liang-Jian, 
  {PMT}
  waveform modeling at the {Daya Bay} experiment, Chinese Physics C 36~(8)
  (2012) 733.

\bibitem{DEAPPMT}
{DEAP-3600 Collaboration}, In-situ characterization methods for the {Hamamatsu}
  5912 photomultiplier tubes used in the {DEAP-3600} experiment~(in
  preparation).

\bibitem{Ap2DChooz}
J.~Haser, F.~Kaether, C.~Langbrandtner, M.~Lindner, S.~Lucht, S.~Roth,
  M.~Schumann, A.~Stahl, A.~Stüken, C.~Wiebusch,
  Afterpulse
  measurements of {R7081} photomultipliers for the {Double Chooz} experiment,
  Journal of Instrumentation 8~(04) (2013) P04029.

\end{thebibliography}

\end{document}